# Icosahedral quasicrystals in Zn-T-Sc (T=Mn, Fe, Co, Ni) alloys

Ryo Maezawa, Shiro Kashimoto and Tsutomu Ishimasa*

*Division of Applied Physics, Graduate School of Engineering, Hokkaido Univ., Kita-ku, Sapporo 060-8628, Japan*

ABSTRACT

Starting from the $Zn_{17}Sc_3$ cubic approximant, new quasicrystal alloys were searched by replacement of Zn with transition elements, T. In the cases of T=Mn, Fe, Co and Ni, new icosahedral quasicrystals are formed in as-cast alloys as major phases at the alloy compositions of $Zn_{75}T_{10}Sc_{15}$. All these quasicrystals belong to a primitive type, and have 6-dimensional lattice parameters, $a_{6D}$, ranging from 7.044 to 7.107 Å. They have the valence electron concentrations, $e/a$, ranging from 2.01 to 2.14, and almost the same ratios between the edge-length of the Penrose tile, $a_R$, and the averaged atomic diameter $\bar{d}$: $a_R/\bar{d}$~1.75. Moreover other Zn- and Cd-based quasicrystals including the same type of atomic cluster, Tsai-type cluster, also have the same values of $e/a$~2.1 and $a_R/\bar{d}$~1.75. The equality in $e/a$ indicates that the Hume-Rothery mechanism plays an important role for the formation of these quasicrystals.

*Corresponding Author

E-mail : ishimasa@eng.hokudai.ac.jp

§ 1. INTRODUCTION

Approximant crystals may be very important and useful to find new quasicrystal alloys. In the framework of the projection method (Duneau and Katz 1985, Elser 1986), a structure of an icosahedral quasicrystal is understood as a projection of a 6-dimensional hypercubic structure onto the 3-dimensional subspace in which the icosahedral symmetry is preserved. An approximant crystal is understood as a rational approximation of the corresponding quasicrystal (Henley 1985, Elser and Henley 1985). In the case of an icosahedral quasicrystal, an approximant includes a local atomic configuration or atomic cluster satisfying the icosahedral symmetry. The Bergman- and Mackay-type clusters are famous examples of these atomic clusters (Elser and Henley 1985, Audier and Guyot 1989). Recently, stimulated by the discovery of binary stable quasicrystals in Cd-Yb and Cd-Ca alloys (Guo et al. 2000a, Tsai et al. 2000), a new type of atomic cluster has been recognized, which we call here 'Tsai-type' cluster. Common characteristics of this cluster is the presence of the triple shells: in the case of the cubic $Zn_{17}Sc_3$ approximant, for example, the cluster consists of the innermost dodecahedral shell of 20 Zn, the icosahedral shell of 12 Sc and the outermost icosidodecahedral shell of 30 Zn (Andrusyak et al. 1989). The Tsai-type cluster can be found in variety of crystalline alloys, namely Be-, Cu-, Zn-, Ga-, Cd- and In-based alloys. Accordingly, there is a lot of possibility to find new quasicrystals with different base metals. Actually, the first Cu-based icosahedral quasicrystal, $Cu_{48}Ga_{34}Mg_3Sc_{15}$ (Kaneko et al. 2002), was discovered by starting from the cubic $Cu_{3.7}Ga_{2.3}Sc$ approximant (Markiv and Belyavina 1983).

The preceding work on Zn-Mg-Sc alloy suggested that the $Zn_{17}Sc_3$ is positioned at a marginal situation to form a quasicrystal: In the case of Zn-Mg-Sc alloy, only 5 at.% substitution of Zn by Mg causes the drastic change from the approximant crystal to the icosahedral quasicrystal (Kaneko et al. 2001). Very recently, it has been clarified that replacement of Zn with noble metals, Pd, Ag, Pt and Au, also causes formation of icosahedral quasicrystals (Kashimoto et al. 2003). The first purpose of the present work is to study an effect of replacement of Zn with transition elements such as Mn, Fe, Co and Ni in the $Zn_{17}Sc_3$ approximant. The second purpose is to obtain general understanding on the formation conditions and the structural properties of this unique type of quasicrystals including Tsai-type cluster.

§ 2. EXPERIMENTAL PROCEDURES

High-purity materials of Zn (Nilaco; purity, 99.998%), Sc (Shin-Etsu Chemical; 99.9%), Mn (Leico Industries, 99.97%), Fe (Nilaco; 99.99%), Co (Hirano-Seizaemon; 99.9%) and Ni (Nilaco; 99.99%) were used to synthesize alloy ingots of nominal composition of $Zn_{75}T_{10}Sc_{15}$, where T denotes Mn, Fe, Co and Ni. Weighed materials were enclosed in a package made of Mo foil (Nilaco, thickness; 0.05 mm; purity, 99.95%) in order to avoid evaporation of Zn and to prevent chemical reaction with the silica ampoule. The specimen was sealed into the silica ampoule in an Ar atmosphere at approximately 260 Torr after previous evacuation to a pressure ranging from 1.2 to $2.0 \times 10^{-6}$ Torr. The specimens were melted at 862±2 ºC for 1 hour, and then cooled in water. In the last process, the specimen was cooled in the silica ampoule through Mo package. Then the cooling speed is relatively slow, and then the specimens are called as-cast ones in this paper. Weight losses during the preparation process were less than 0.7 %.

Structure characterization was carried out by means of powder X-ray diffraction, transmission electron microscopy and scanning electron microscopy using wavelength-dispersive X-ray spectroscopy; *i.e.* electron probe microanalysis, EPMA. Composition analysis by the EPMA method was carried out after detection of the domain of each phase by observing a back-scattered electron image. An average value of the composition of each alloy was estimated from more than 4, typically 9, measurements at different parts of the samples. The spatial fluctuation of the composition was less than 1 at.% in each phase. No Mo contamination from the package was confirmed by the EPMA analysis. Selected-area electron diffraction experiments were carried out with a JEOL 200CS microscope operated at an acceleration voltage of 120 kV. Other details of the experimental procedures were described elsewhere (Kaneko at al. 2001).

## § 3. RESULTS AND DISCUSSION

Formation of primitive-type (P-type) icosahedral quasicrystals were detected in the as-cast $Zn_{75}T_{10}Sc_{15}$ alloys, where T = Mn, Fe, Co and Ni. The 2-, 3-, and 5-fold diffraction patterns of the Zn-Mn-Sc icosahedral quasicrystal are presented in figures 1 (a), (b) and (c), respectively, and those of the Zn-Ni-Sc in figures 1(d) ~ (f). The diffraction symmetry, $m\overline{35}$, clearly indicates the formation of the icosahedral quasicrystals in both alloys. All the spots observed in these patterns can be indexed using six integers (Elser 1986). There is no reflection condition in both quasicrystals. This observation indicates the formation of P-type icosahedral quasicrystals in these alloys. The $\tau^3$ scaling characteristic of the P-type can be noticed, for example, along the 5-fold axis in figure 1 (a). Here $\tau$ denotes the golden mean, $(1+\sqrt{5})/2$. No reflections corresponding to $\tau$ or $\tau^2$ scaling are observed in figures 1 (a) and (d). In other alloys with T=Fe and Co, the existence of P-type icosahedral quasicrystals were also confirmed by electron microscopic observations.

In figure 2, powder X-ray diffraction patterns of the four alloys are presented, which were measured by Cu K$\alpha$ radiation. The analysis of these patterns confirms the formation of the P-type icosahedral quasicrystals in these alloys as major phases. In figure 2 (d), diffraction peaks due to the Zn-Ni-Sc quasicrystal are indicated by six indices according to the indexing scheme proposed by Elser (1986). The complete list of the observed reflections is presented in table 1, in which in the fifth column the indices of the reflections are listed. There is no special restriction on the indices in accordance with the observation of the electron diffraction patterns. Therefore the formation of the P-type icosahedral quasicrystal was confirmed in the as-cast $Zn_{75}Ni_{10}Sc_{15}$ alloy. The six-dimensional lattice parameter was estimated to be $a_{6D}$=7.044±0.001 Å using the extrapolation method described in the former letter (Kaneko et al. 2001). The lattice spacings of the reflections calculated from the lattice parameter are listed in the fourth column of table 1. They agree well with the measured values, $d_{ob}$, listed in the third column of table 1. The intensity distribution in the diffraction pattern in figure 2 (d) is very similar to those of the $Zn_{81}Mg_4Sc_{15}$ and $Cu_{48}Ga_{34}Mg_3Sc_{15}$ quasicrystals: See figures in the articles by Ishimasa et al. (2002) and Kaneko et al. (2002). The composition of the icosahedral quasicrystal was estimated to be $Zn_{74}Ni_{10}Sc_{16}$ by the EPMA measurement. This analyzed composition is approximately equal to the nominal composition $Zn_{75}Ni_{10}Sc_{15}$. The majority of the specimen is occupied by the icosahedral quasicrystal, but the EPMA analysis indicated presence of an impurity phase with an alloy composition of $Zn_{74}Ni_{17}Sc_9$. This impurity phase may be related to the unidentified reflections indicated by asterisks in table 1.

The formation of the P-type icosahedral quasicrystals were also confirmed in the alloys with T=Mn, Fe, and Co by powder X-ray diffraction experiments. In particular, in the cases of the $Zn_{75}Mn_{10}Sc_{15}$ and $Zn_{75}Fe_{10}Sc_{15}$ alloys, the icosahedral quasicrystals are formed almost exclusively. The composition of the Zn-Mn-Sc icosahedral quasicrystal was analyzed to be $Zn_{74}Mn_{10}Sc_{16}$ by the EPMA method, which is nearly equal to the nominal composition. This analyzed composition corresponds to that of Zn-Ni-Sc quasicrystal. Accordingly, there may be a stoichiometric composition for this new series of icosahedral quasicrystals. The six-dimensional lattice parameters, $a_{6D}$, were measured to be 7.107±0.001 Å, 7.083±0.001 Å, and 7.061±0.001 Å, for the quasicrystals formed in $Zn_{75}Mn_{10}Sc_{15}$, $Zn_{75}Fe_{10}Sc_{15}$ and $Zn_{75}Co_{10}Sc_{15}$ alloys, respectively.

The six-dimensional lattice parameters, $a_{6D}$, of these quasicrystals decrease gradually from 7.107 to 7.044 Å with the increase of atomic numbers of the transition elements. This may be related to the decrease of atomic sizes of the transition elements. Similar decrease in $a_{6D}$ was also recognized by Guo et al. (2000b) in the series of the icosahedral quasicrystals, $Cd_{65}Mg_{20}L_{15}$, where L denotes the lanthanoid metals such as Gd, Tb, Dy, Ho, Er, Tm, Yb and Lu. The $Cd_{65}Mg_{20}L_{15}$ icosahedral quasicrystals are expected to include Tsai-type cluster from the structure of the corresponding approximant, $Cd_6Yb$ (Palenzona 1971). Guo et al. (2000b) noticed that the edge-lengths of Penrose tile, $a_R=a_{6D}/\sqrt{2}$, are proportional to the averaged atomic diameters, $\bar{d}$, in the case of the Cd-based quasicrystals: $a_R \approx 1.75\bar{d}$. The ratio, $a_R/\bar{d}$, may reflect the atomic decoration of the Penrose rhombohedra, and thus reflects the type of atomic cluster included in each quasicrystal. For the Mackay-type quasicrystals such as F-$Al_{70}Pd_{21}Mn_9$ and F-$Al_{65}Cu_{21}Fe_{14}$, this ratio is approximately 1.62. Here a half of $a_F$, the six-dimensional lattice parameter of the F lattice, is used as the six-dimensional lattice parameter for comparison. In the case of Bergman type, this ratio ranges from 1.72 for $Zn_{40}Mg_{40}Al_{20}$ to 1.74 for $Zn_{60}Mg_{30}Ho_{10}$. The values of $a_R$ and $\bar{d}$ are summarized in table 2 for the icosahedral quasicrystals found in the past three years, all of which may be classified into Tsai-type. This table also includes the data of the $Zn_{75}M_{10}Sc_{15}$, M=Pd, Ag, Pt and Au (Kashimoto et al. 2003). They have close relation to the present series with the same stoichiometry, but with the different kind of metals, namely noble metals. The ratios, $a_R/\bar{d}$, range between 1.73 to 1.77 for all the quasicrystals listed in table 2. This agreement is related to the fact that all these quasicrystals belong to the unique structural type, *i.e.* Tsai-type, even though the sizes of the Penrose rhombohedra distributed rather widely from 4.906 to 5.731 Å. In figure 3, the values of $a_R$ are proportional to those of $\bar{d}$ with the relation, $a_R=1.75\bar{d}$. In this diagram, the Zn-M-Sc and Zn-T-Sc quasicrystals are located in the restricted region indicated by the ellipsoid. This fact indicates that isomorphs substitution is realized not only for the transition elements but also for the noble metals, Pd, Ag, Pt and Au. Therefore, it is concluded that the transition elements, Mn, Fe, Co and Ni, and the noble metals play the same role to stabilize this unique type of icosahedral quasicrystal. It is noted that the Bergman and Tsai-types have almost the same magnitude of $a_R/\bar{d}$. However, for the present moment, we do not know the reason of this coincidence.

The Hume-Rothery mechanism due to the Brillouin zone and the Fermi surface interaction (Friedel 1988, Tsai 1998) and the sp-d hybridization (Takeuchi et al. 2002, Ishii and Fujiwara 2002) have been discussed frequently for the stabilization of a quasicrystal. In the present case of the Zn-T-Sc alloys, it is difficult to evaluate the valence electron concentration, $e/a$, because of the presence of the transition elements. It is well-known that the valence of a transition element depends on alloying components ( Elliott and Rostoker 1958, Sinha 1973). For the calculation of $e/a$ in table 2, we tentatively used the valences proposed by Haworth and Hume-Rothery (1952): Mn; 1.9, Fe; 1.0, Co; 0.8 and Ni; 0.6, which explain the formation of the $\alpha/\beta$ Brass type in Cu-Zn(Al)-T alloys. For the noble metals, the valences are assumed to be 0 for Pd and Pt, and 1 for Ag and Au. The values of $e/a$ are approximately equal to each other for the new Zn-T-Sc quasicrystals: They range from 2.01 to 2.14. Furthermore these values are approximately equal to those of other quasicrystals including Zn-M-Sc as listed in table 2. This approximate equality of $e/a$ indicates the role of Hume-Rothery mechanism in the stabilization of these quasicrystals. It should be noted that no such equality is found, if we use the valences proposed by Raynor (1949): Mn; -3.66, Fe; -2.66, Co; -1.71 and Ni; -0.61 for Al-rich intermediate phases, which explain beautifully the formation of Al-transition element quasicrystals (Tsai 1998). The large difference in the appropriate valences is not special for the quasicrystals, but common for crystalline intermetallic phases (Elliott and Rostoker 1958). It should be added that the discovery of Zn-T(M)-Sc quasicrystals is also favorable for the sp-d hybridization mechanism.

## § 4. CONCLUSION

In the as-cast alloys of $Zn_{75}T_{10}Sc_{15}$ (T=Mn, Fe, Co and Ni), the P-type icosahedral quasicrystals are formed as major phases. This result supports the interpretation that the $Zn_{17}Sc_3$ approximant is situated at a marginal condition for the formation of an icosahedral quasicrystal. The icosahedral cluster, Tsai-type cluster, included in the $Zn_{17}Sc_3$ approximant seems energetically stable against the additives of the transition elements. Furthermore they destroy the periodic order of the approximant crystal and create new quasiperiodic order. The Hume-Rothery rule seems valid for the newly created quasicrystals. Accordingly, the systematic replacement in the $Zn_{17}Sc_3$ approximant may give us important hint for understanding the factors which govern the existence of a quasicrystal.

## Acknowledgments

The authors wish to thank Mr. N. Miyazaki for the use of an electron probe microanalyzer JXA-8900M. This work was supported by a Grant-in-Aid for Scientific Research B from Japan Society for the Promotion of Science.

Figure Captions

Figure 1. Electron diffraction patterns of the Zn-Mn-Sc and Zn-Ni-Sc quasicrystals. Upper and lower three patterns correspond to the Zn-Mn-Sc and Zn-Ni-Sc, respectively. (a) and (d): along two-fold , (b) and (e): three-fold, and (c) and (f): five-fold axes. The arrowhead in (a) indicates the $01\bar{1}\bar{2}02$ reflection with $d$=2.257 Å.

Figure 2. Powder X-ray diffraction patterns of $Zn_{75}T_{10}Sc_{15}$ alloys measured by Cu Kα radiation. (a) T=Mn, (b) T=Fe, (c) T=Co, and (d) T=Ni. Arrows in (d) indicate unindexed reflections due to an impurity phase.

Figure 3. Relation between the edge-length of the Penrose tile, $a_R$, and the averaged atomic diameter, $\bar{d}$.

Table 1. Results of powder X-ray diffraction experiment of $Zn_{75}Ni_{10}Sc_{15}$ alloy. The five columns correspond to observed 2θ values, the peak intensity, the measured *d* values, the calulated *d* values, and the indices of the quasicrystal reflection. The 6-dimensional lattice parameter, $a_{6D}$ = 7.044 Å, was used for $d_{cal}$. The asterisks in the fourth column indicate reflections due to an impurity phase.

| 2θ (degree) | Intensity | $d_{ob}$ (Å) | $d_{cal}$ (Å) | Indices |
|---|---|---|---|---|
| 8.82 | 1 | 10.02 | 9.962 | $0\ 0\ 0\ 0\ 0\ 1$ |
| 15.06 | 1 | 5.88 | 5.855 | $0\ 0\ 0\ \bar{1}\ 0\ 1$ |
| 21.20 | 7 | 4.187 | 4.179 | $0\ 1\ 0\ \bar{1}\ 0\ 1$ |
| 24.54 | 2 | 3.624 | 3.619 | $0\ 1\ \bar{1}\ \bar{1}\ 0\ 1$ |
| 26.22 | 1 | 3.396 | 3.401 | $1\ 1\ \bar{1}\ \bar{1}\ 0\ 1$ |
| 28.96 | 2 | 3.080 | 3.078 | $0\ 1\ 0\ \bar{1}\ 0\ 2$, $1\ 1\ \bar{1}\ \bar{1}\ 1\ 1$ |
| 30.34 | 2 | 2.943 | 2.941 | $0\ 1\ 0\ \bar{1}\ 1\ 2$, $0\ 2\ \bar{1}\ \bar{1}\ 0\ 1$ |
| 32.70 | 1 | 2.736 | 2.736 | $0\ 1\ \bar{1}\ \bar{1}\ 0\ 2$ |
| 34.02 | 1 | 2.633 | 2.628 | $1\ 1\ \bar{1}\ \bar{2}\ 1\ 1$, $1\ 1\ \bar{1}\ \bar{1}\ \bar{1}\ 2$ |
| 34.82 | 1 | 2.574 | * | |
| 38.20 | 42 | 2.354 | 2.352 | $1\ 1\ \bar{1}\ \bar{1}\ 1\ 2$ |
| 39.42 | 1 | 2.284 | 2.282 | $1\ 2\ \bar{1}\ \bar{2}\ 0\ 1$, $0\ 2\ \bar{1}\ \bar{1}\ 1\ 2$ |
| 40.24 | 100 | 2.239 | 2.237 | $0\ 1\ \bar{1}\ \bar{2}\ 0\ 2$ |
| 41.26 | 1 | 2.186 | * | |
| 41.36 | 2 | 2.181 | 2.182 | $1\ 1\ \bar{1}\ \bar{2}\ 0\ 2$ |
| 43.04 | 14 | 2.100 | * | |
| 44.22 | 1 | 2.046 | 2.045 | $1\ 1\ \bar{1}\ \bar{1}\ 0\ 3$ |
| 45.94 | 39 | 1.974 | 1.972 | $0\ 2\ \bar{1}\ \bar{2}\ 0\ 2$ |
| 47.00 | 1 | 1.932 | 1.931 | $0\ 2\ \bar{1}\ \bar{1}\ 0\ 3$, $1\ 2\ \bar{1}\ \bar{2}\ 1\ 2$ |
| 47.72 | 14 | 1.904 | 1.903 | $1\ 1\ \bar{1}\ \bar{1}\ 1\ 3$, $1\ 2\ \bar{1}\ \bar{2}\ 0\ 2$ |
| 48.66 | 2 | 1.870 | 1.869 | $0\ 1\ \bar{1}\ \bar{2}\ 0\ 3$, $1\ 2\ \bar{1}\ \bar{2}\ \bar{1}\ 2$, $1\ 2\ \bar{1}\ \bar{2}\ 1\ 2$ |
| 50.34 | 1 | 1.811 | 1.809 | $0\ 2\ \bar{2}\ \bar{2}\ 0\ 2$, $1\ 1\ \bar{1}\ \bar{2}\ 0\ 3$ |
| 51.22 | 1 | 1.782 | 1.780 | $0\ 2\ 0\ \bar{2}\ 0\ 3$ |
| 52.30 | 1 | 1.748 | * | |
| 56.72 | 1 | 1.622 | 1.621 | $1\ 2\ \bar{1}\ \bar{2}\ 0\ 3$ |
| 58.22 | 1 | 1.583 | 1.581 | $1\ 2\ \bar{1}\ \bar{2}\ 1\ 3$ |
| 60.52 | 1 | 1.529 | 1.527 | $1\ 2\ \bar{2}\ \bar{2}\ 0\ 3$ |
| 61.44 | 1 | 1.508 | 1.509 | $0\ 3\ \bar{1}\ \bar{2}\ 0\ 3$, $1\ 2\ \bar{1}\ \bar{2}\ 2\ 3$ |
| 62.46 | 1 | 1.486 | * | |
| 62.76 | 1 | 1.479 | 1.479 | $0\ 2\ \bar{1}\ \bar{3}\ 0\ 3$, $1\ 2\ \bar{2}\ \bar{2}\ 1\ 3$ |
| 64.14 | 2 | 1.451 | 1.449 | $1\ 2\ \bar{1}\ \bar{3}\ 0\ 3$ |
| 64.90 | 1 | 1.436 | 1.434 | $1\ 2\ \bar{1}\ \bar{3}\ 1\ 3$, $1\ 2\ \bar{1}\ \bar{3}\ 1\ 3$ |
| 67.66 | 12 | 1.384 | 1.382 | $0\ 2\ \bar{2}\ \bar{3}\ 0\ 3$ |
| 68.38 | 2 | 1.371 | 1.369 | $1\ 2\ \bar{2}\ \bar{3}\ 0\ 3$, $1\ 2\ \bar{1}\ \bar{2}\ 1\ 4$ |
| 69.76 | 1 | 1.347 | * | |
| 71.82 | 5 | 1.313 | 1.312 | $1\ 3\ \bar{1}\ \bar{3}\ 0\ 3$ |
| 73.16 | 5 | 1.292 | 1.291 | $1\ 2\ \bar{2}\ \bar{2}\ 1\ 4$, $1\ 3\ \bar{1}\ \bar{3}\ \bar{1}\ 3$ |
| 75.38 | 1 | 1.260 | 1.261 | $1\ 2\ \bar{1}\ \bar{3}\ 1\ 4$, $1\ 3\ \bar{2}\ \bar{3}\ 0\ 3$ |
| 77.20 | 3 | 1.235 | 1.234 | $0\ 2\ \bar{2}\ \bar{3}\ 0\ 4$, $2\ 2\ \bar{2}\ \bar{2}\ 1\ 4$ |
| 78.90 | 1 | 1.212 | 1.216 | $1\ 2\ \bar{2}\ \bar{3}\ 0\ 4$ |
| 81.78 | 2 | 1.177 | 1.176 | $1\ 3\ \bar{1}\ \bar{3}\ 0\ 4$, $2\ 2\ \bar{2}\ \bar{2}\ 2\ 4$ |

Table 2. Summary of Tsai-type icosahedral quasicrystals. For the calulation of $\bar{d}$, atomic radii in the article by Pearson (1972) are used. The following valencies were used for the calculation of $e/a$: 1 for Cu, 2 for Mg, Ca, Zn, Cd and Yb, 3 for Sc, Ga, Y, In, Gd, Tb, Dy, Ho, Er, Tm and Lu, and 4 for Ti, respectively. For Mn, Fe, Co, Ni, Pd, Ag, Pt and Au see text.

| Composition | $a_R$ (Å) | $\bar{d}$ (Å) | $a_R/\bar{d}$ | $e/a$ | References |
|---|---|---|---|---|---|
| $Cu_{48}Ga_{34}Mg_3Sc_{15}$ | 4.906 | 2.775 | 1.77 | 2.01 | Kaneko et al. (2002) |
| $Zn_{84}Mg_8Ti_8$ | 4.966 | 2.832 | 1.75 | 2.16 | Ishimasa et al. (2002) |
| $Zn_{75}Ni_{10}Sc_{15}$ | 4.981 | 2.833 | 1.76 | 2.01 | Present work |
| $Zn_{75}Co_{10}Sc_{15}$ | 4.993 | 2.834 | 1.76 | 2.03 | Present work |
| $Zn_{75}Fe_{10}Sc_{15}$ | 5.008 | 2.838 | 1.76 | 2.05 | Present work |
| $Zn_{75}Mn_{10}Sc_{15}$ | 5.025 | 2.844 | 1.77 | 2.14 | Present work |
| $Zn_{80}Mg_5Sc_{15}$ | 5.028 | 2.883 | 1.74 | 2.15 | Kaneko et al. (2001) |
| $Zn_{75}Pd_{10}Sc_{15}$ | 5.030 | 2.859 | 1.76 | 1.95 | Kashimoto et al. (2003) |
| $Zn_{75}Pt_{10}Sc_{15}$ | 5.029 | 2.861 | 1.76 | 1.95 | ibid. |
| $Zn_{75}Ag_{10}Sc_{15}$ | 5.054 | 2.872 | 1.76 | 2.05 | ibid. |
| $Zn_{75}Au_{10}Sc_{15}$ | 5.057 | 2.872 | 1.76 | 2.05 | ibid. |
| $Zn_{75}Mg_{14}Er_{11}$ | 5.130 | 2.926 | 1.75 | 2.11 | Sterzel et al. (2002) |
| $Zn_{76}Mg_{10}Yb_{14}$ | 5.211 | 2.982 | 1.75 | 2.00 | Mitani et al. (2003) |
| $Ag_{42}In_{42}Yb_{16}$ | 5.590 | 3.232 | 1.73 | 2.00 | Guo et al. (2002) |
| $Ag_{42}In_{42}Ca_{16}$ | 5.606 | 3.242 | 1.73 | 2.00 | ibid. |
| $Cd_{65}Mg_{20}Lu_{15}$ | 5.571 | 3.199 | 1.74 | 2.15 | Guo et al. (2000b) |
| $Cd_{65}Mg_{20}Tm_{15}$ | 5.606 | 3.203 | 1.75 | 2.15 | ibid. |
| $Cd_{65}Mg_{20}Y_{15}$ | 5.606 | 3.220 | 1.74 | 2.15 | ibid. |
| $Cd_{65}Mg_{20}Er_{15}$ | 5.622 | 3.206 | 1.75 | 2.15 | ibid. |
| $Cd_{65}Mg_{20}Ho_{15}$ | 5.625 | 3.209 | 1.75 | 2.15 | ibid. |
| $Cd_{65}Mg_{20}Dy_{15}$ | 5.628 | 3.211 | 1.75 | 2.15 | ibid. |
| $Cd_{65}Mg_{20}Tb_{15}$ | 5.628 | 3.214 | 1.75 | 2.15 | ibid. |
| $Cd_{65}Mg_{20}Gd_{15}$ | 5.648 | 3.220 | 1.75 | 2.15 | ibid. |
| $Cd_{65}Mg_{20}Yb_{15}$ | 5.727 | 3.261 | 1.76 | 2.00 | ibid. |
| $Cd_{65}Mg_{20}Ca_{15}$ | 5.731 | 3.271 | 1.75 | 2.00 | ibid. |
| $Cd_{84}Yb_{16}$ | 5.697 | 3.255 | 1.75 | 2.00 | Guo et al. (2000a) |
| $Cd_{85}Ca_{15}$ | 5.731 | 3.258 | 1.76 | 2.00 | ibid. |

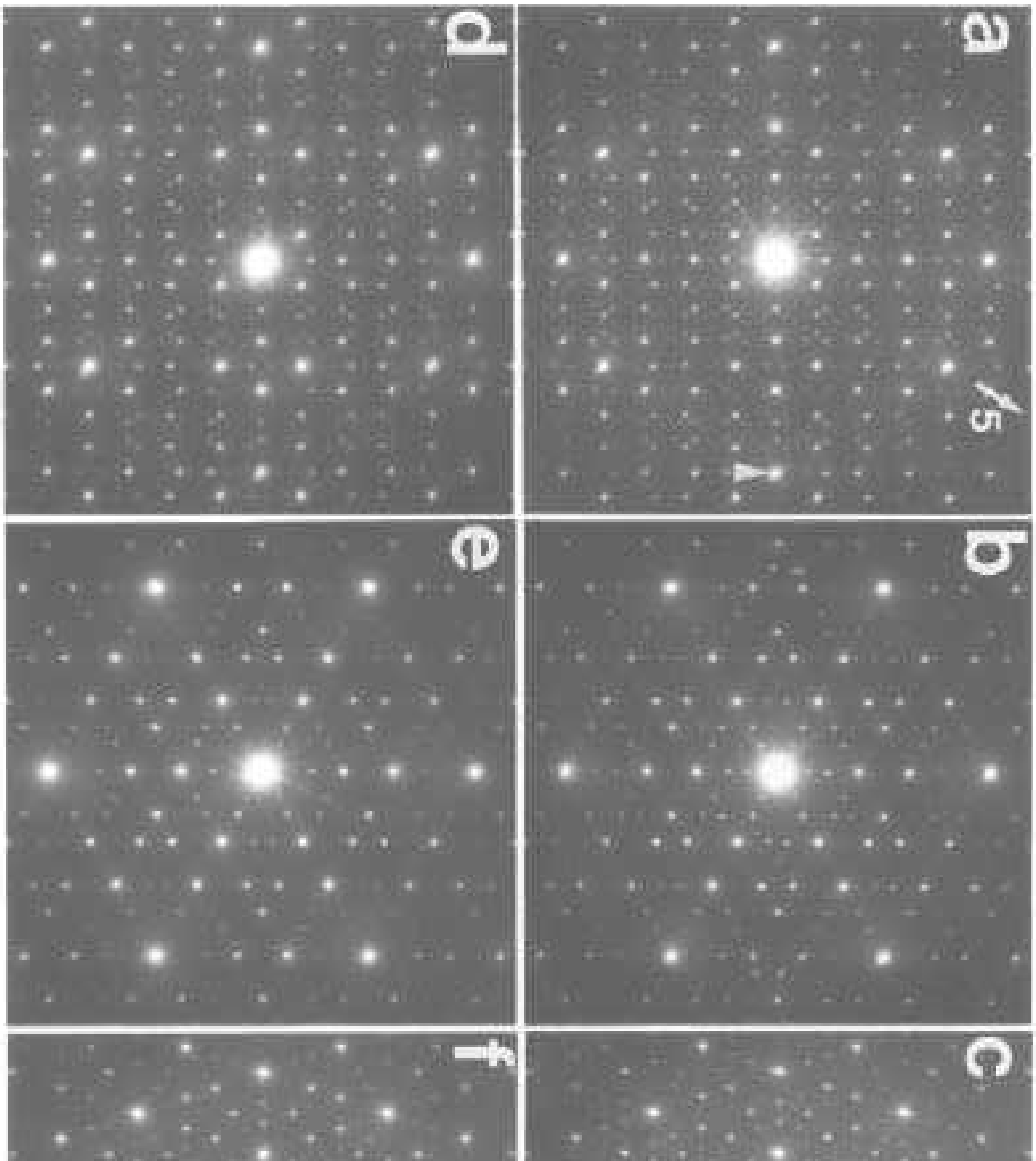

Fig.1

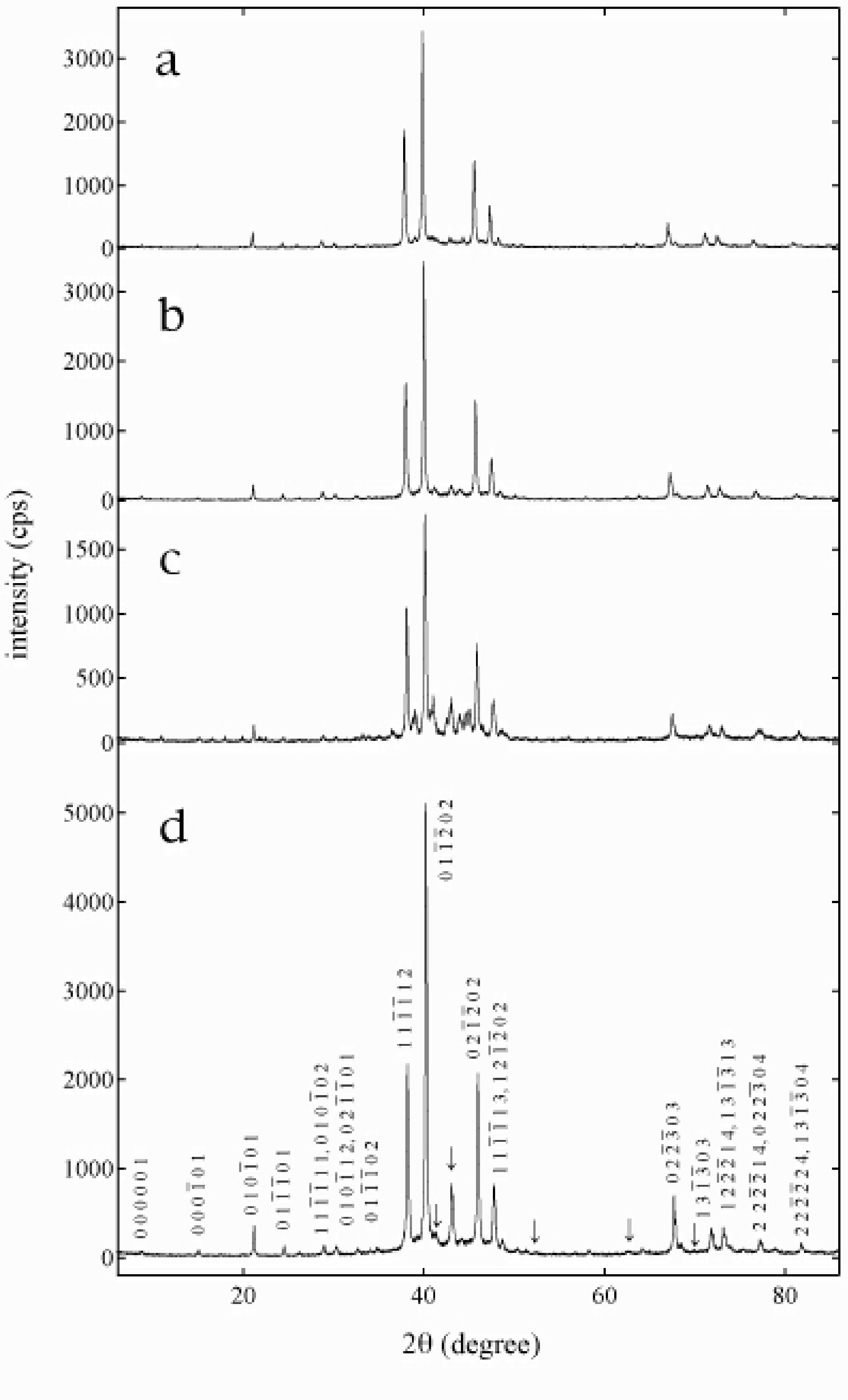

Fig. 2

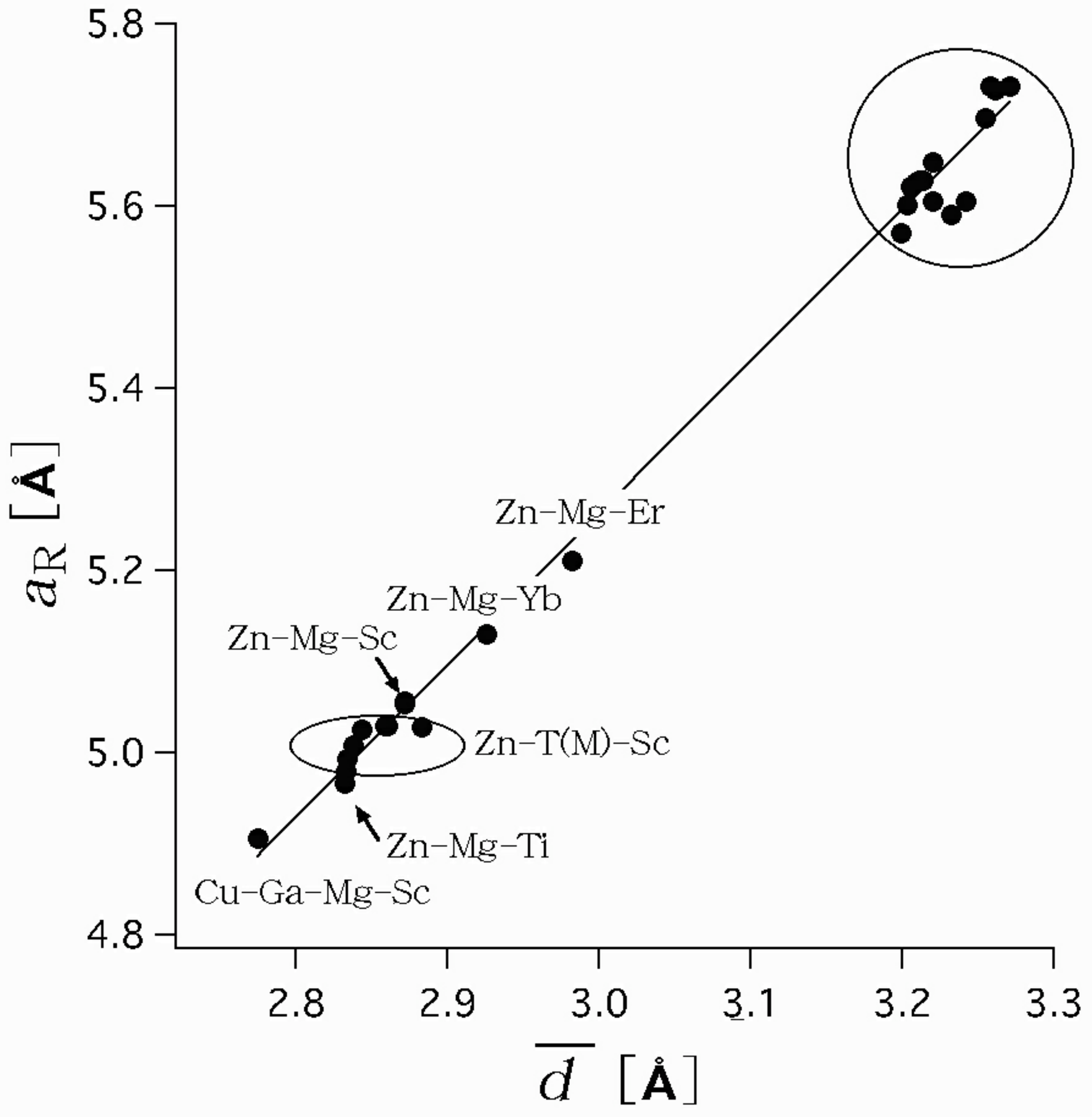

Fig. 3 (Maezawa et al.)